\documentclass{webofc}
\usepackage[varg]{txfonts}
\usepackage{hyperref}
\usepackage{booktabs}
\usepackage[nolist,nohyperlinks]{acronym}
\usepackage{xspace}
\acrodef{AGC}{Analysis Grand Challenge}
\acrodef{ML}{Machine Learning}
\acrodef{BDT}{Boosted Decision Tree}
\newcommand{\ttbar}{\ensuremath{t\bar{t}}\xspace}
\begin{document}
\title{Physics analysis for the HL-LHC: concepts and pipelines in practice with the Analysis Grand Challenge}
\author{
    \firstname{Alexander} \lastname{Held}\inst{1}\fnsep\thanks{\email{alexander.held@cern.ch}} \and
    \firstname{Elliott} \lastname{Kauffman}\inst{2} \and
    \firstname{Oksana} \lastname{Shadura}\inst{3} \and
    \firstname{Andrew} \lastname{Wightman}\inst{3}
}
\institute{
    University of Wisconsin–Madison, United States \and
    Princeton University, United States \and
    University of Nebraska–Lincoln, United States
}
\abstract{%
Realistic environments for prototyping, studying and improving analysis workflows are a crucial element on the way towards user-friendly physics analysis at HL-LHC scale.
The IRIS-HEP \ac{AGC} provides such an environment.
It defines a scalable and modular analysis task that captures relevant workflow aspects, ranging from large-scale data processing and handling of systematic uncertainties to statistical inference and analysis preservation.
By being based on publicly available Open Data, the \ac{AGC} provides a point of contact for the broader community.
Multiple different implementations of the analysis task that make use of various pipelines and software stacks already exist.

This contribution presents an updated \ac{AGC} analysis task.
It features a machine learning component and expanded analysis complexity, including the handling of an extended and more realistic set of systematic uncertainties.
These changes both align the \ac{AGC} further with analysis needs at the HL-LHC and allow for probing an increased set of functionality.

Another focus is the showcase of a reference \ac{AGC} implementation, which is heavily based on the HEP Python ecosystem and uses modern analysis facilities.
The integration of various data delivery strategies is described, resulting in multiple analysis pipelines that are compared to each other.
}
\maketitle
\acresetall
\section{Introduction}

IRIS-HEP~\cite{irishep} is the Institute for Research and Innovation in Software for High Energy Physics.
It is comprised of physicists, computer scientists, and engineers from institutes across the United States who perform research and development to address HL-LHC computing challenges by developing suitable software cyberinfrastructure.
Work in the institute proceeds in various topical areas, including data reconstruction and triggering in the innovative algorithms area, the analysis systems area aiming to reduce time-to-insight and maximize physics potential, as well as the data organization, management, and access systems area.

A variety of new ideas for implementing physics analyses in the future are being pursued in IRIS-HEP and the broader community, alongside new tools and techniques which are being developed.
The IRIS-HEP \ac{AGC} project~\cite{Held:2022sfw} started in this context as an integration exercise: combine the ongoing developments to demonstrate an efficient end-to-end analysis pipeline that is designed for use at the HL-LHC.
There are two aspects to the \ac{AGC}:
\begin{itemize}
    \item define a physics analysis task of realistic scope and scale, representative of HL-LHC requirements,
    \item develop analysis pipelines that implement the task, find and address performance bottlenecks and usability concerns in the process.
\end{itemize}

\subsection{Goals of the AGC project}

Despite starting out as just an integration exercise, over time it became clear that the \ac{AGC} project can be useful in additional ways to the broader community.
For developers of software and infrastructure components, it provides a realistic testbed for evaluating user experience, interface design and performance.
It is an example workload for testing modern analysis facilities and new services, allowing to explore the possibility of "interactive" analysis.
Achieving an interactive turnaround time of minutes or less requires highly parallel execution with low latency, which also mandates efficient use of intermediate caching.

The \ac{AGC} project has organized a series of dedicated workshops~\cite{agc-tools-2021,agc-tools-2022} to showcase related work and demonstrate the envisioned analysis pipelines.
These workshops provide dedicated opportunities to interact with the broader community and receive valuable input regarding the envisioned approaches showcased.

\section{The AGC analysis task}

The \ac{AGC} focuses on the last pieces in a physics analysis pipeline, starting out from centrally produced common data samples.
To capture the needs of a physicist in their practical applications, the analysis task must probe the following workflow aspects:
\begin{itemize}
    \item extracting and filtering data from common data samples,
    \item calibrating objects and evaluating systematic variations,
    \item \ac{ML} training and inference,
    \item statistical model building and inference,
    \item relevant visualizations.
\end{itemize}

A top quark pair production (\ttbar) cross-section measurement in the single lepton channel is chosen for the \ac{AGC} as analysis task.
It uses CMS Open Data from 2015~\cite{cms-open-data}, allowing participation from anyone interested without the need for special permissions to access this data.
This Open Data release is available in MiniAOD~\cite{Petrucciani:2015gjw} format.
In order to more closely correspond to the data formats expected to be used at the HL-LHC (PHYSLITE~\cite{physlite} for ATLAS, NanoAOD~\cite{nanoaod} for CMS), the relevant MiniAOD files are pre-converted to NanoAOD format.
These converted files are available publicly in XRootD-accessible storage at the University of Nebraska–Lincoln.
Details of where to find them are available in the \ac{AGC} GitHub repository~\cite{agc_code}.
In total, about 1 billion events are used, corresponding to a file size of around 2 TB.
The \ac{AGC} analysis task, as well as implementations of it, are strictly aimed at demonstrating workflows and functionality instead of physics results.

The analysis task is set up to allow for probing the three biggest user experience pain points in physics analysis as highlighted in the Second Analysis Ecosystem Workshop report~\cite{stewart_graeme_a_2022_7003963}: dealing with systematic uncertainties, handling metadata, and scaling a pipeline from a prototype to a sufficiently powerful facility.

A new addition for this conference to the \ac{AGC} analysis task is a \ac{ML} component.
This was frequently requested by the community, owing to the ubiquitous use of \ac{ML} in physics analysis.
For the \ac{AGC}, the \ac{ML} task is the correct matching of reconstructed objects to constituents in the decay of the \ttbar system.
In practice, this implies the need to enhance the analysis pipeline: adding a \ac{ML} model training step, as well as subsequent \ac{ML} inference.
It also changes the computational needs of the analysis pipeline: complex \ac{ML} models may need dedicated specialized resources for efficient training and possibly inference.

\subsection{Analysis task versions}

With the \ac{AGC} analysis task expanding, a versioning scheme for analysis tasks allows to keep track of what each task corresponds to.
Table~\ref{agc-task-versions} lists the three major versions that exist.
Version 0 is fully superseded.
Version 1 provides a stable baseline before the addition of \ac{ML} to the pipeline.
Work towards finalizing version 2 is ongoing, this version includes both \ac{ML} aspects and additional computational complexity via an enhanced set of systematic uncertainties.
The \ac{AGC} documentation webpage~\cite{agc_rtd} lists additional details about the analysis task definition.
\begin{table}
    \centering
    \caption{Versions of the \ac{AGC} analysis task.}
    \label{agc-task-versions}
    \begin{tabular}{ll}
        \toprule
        version & description  \\
        \midrule
        v0 & input files in custom ntuple format \\
        v1 & input files in CMS NanoAOD format, minimal changes to analysis logic \\
        v2 (planned) & addition of \ac{ML}, expanded set of systematic uncertainties \\
        \bottomrule
    \end{tabular}
\end{table}

\section{The AGC reference implementation}

\begin{figure}[ht]
    \centering
    \includegraphics[width=0.95\linewidth]{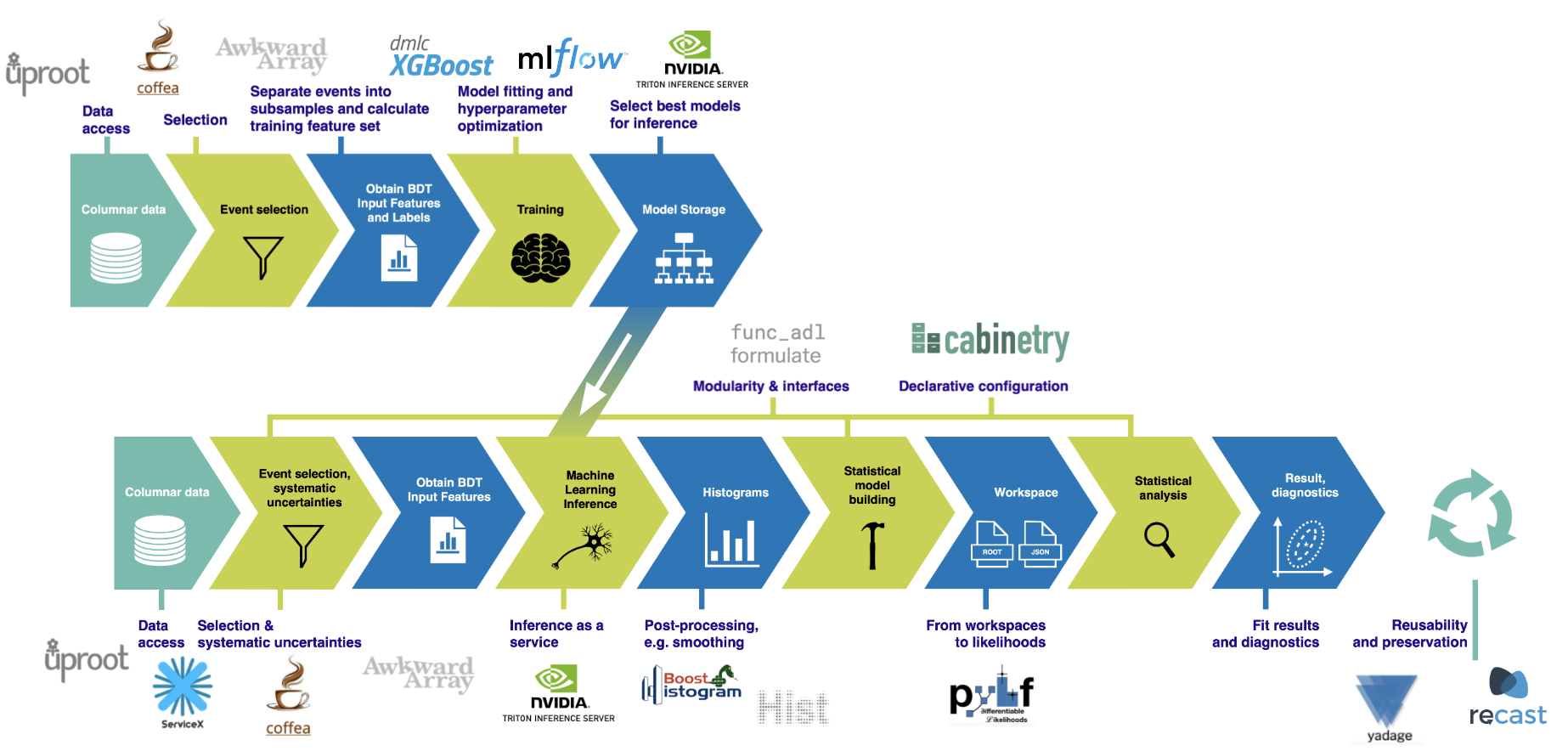}
    \caption{The reference implementation for the \ac{AGC} analysis task.}
    \label{agc-pipeline}
\end{figure}

IRIS-HEP provides a Python-based reference implementation for the AGC analysis task on GitHub~\cite{agc_code}.
It uses tools and services developed within IRIS-HEP and the community, including many Scikit-HEP~\cite{Rodrigues:2020syo} libraries.
The repository also includes information about the datasets that are being used in the \ac{AGC}.
Figure~\ref{agc-pipeline} schematically shows the pipeline and how it is implemented in practice.

The pipeline starts with an optional component: \textit{ServiceX}~\cite{servicex}, a data delivery service.
After receiving a declarative request via \textit{FuncADL}~\cite{funcadl}, \textit{ServiceX} can perform filtering and return the required columns for subsequent processing.
In the implementation discussed here, the output of \textit{ServiceX} is written to object storage at an analysis facility, which serves as a cache.
Subsequent executions of the analysis pipeline (without changes in the request sent to \textit{ServiceX}) will be able to make use of this cache to speed up turnaround time.
\textit{ServiceX} can be co-located with the data it ingests for optimal performance, limiting the amount of data that needs to be transferred after the initial filtering that can already take place within \textit{ServiceX}.

The next parts of the pipeline are steered via the \textit{coffea}~\cite{coffea} framework, which controls the distributed event processing and the production of histograms.
Distribution is handled in this implementation by \textit{Dask}~\cite{dask}, but \textit{coffea} supports additional alternatives.
When \textit{ServiceX} is used, \textit{coffea} reads input files in ROOT format~\cite{Brun:1997pa} out of the object storage via \textit{uproot}~\cite{uproot}; without \textit{ServiceX} it instead streams the remote NanoAOD files over the network.
Within the code executed via \textit{coffea}, many other libraries from the Scikit-HEP project provide functionality that is being used here.
The \textit{awkward-array}~\cite{awkward} library enables efficient columnar data processing, \textit{boost-histogram}~\cite{boost-histogram} and \textit{hist}~\cite{hist} are used to produce histograms from columnar data.
Systematic variations are applied to data via \textit{correctionlib}~\cite{correctionlib}, which is another new addition for this conference.
It provides a standardized JSON format for defining corrections and an interface for applying them.
\ac{ML} inference can optionally be outsourced via a \textit{NVIDIA Triton}~\cite{NVIDIA_Corporation_Triton_Inference_Server} inference server to be efficiently executed on dedicated resources.
Following the production of all required histograms, \textit{cabinetry}~\cite{cabinetry} constructs a statistical model.
Statistical inference is peformed using \textit{pyhf}~\cite{pyhf,pyhf_joss}.
Additional visualization utilities are provided by \textit{mplhep}~\cite{mplhep}.

In addition to this main processing pipeline, the top part of figure~\ref{agc-pipeline} shows the \ac{ML} training part.
It uses a similar approach to prepare the data required for training, with \textit{coffea}, \textit{uproot} and \textit{awkward-array} playing major roles.
A \ac{BDT} is trained using \textit{XGBoost}~\cite{xgboost}.
\textit{MLflow}~\cite{mlflow} provides functionality for experiment tracking to store performance metrics of different models during hyperparameter optimization.
The best-performing models are uploaded for use via \textit{NVIDIA Triton}.

\subsection{Reference implementation versions}

Several versions of the \ac{AGC} reference implementation exist.
In the versioning scheme used, the major version corresponds to the version of the analysis task as shown in table~\ref{agc-task-versions}.
The first available version, v0.1, is used for the benchmarking results presented at the ACAT 2022 conference~\cite{acat_proceedings}.
Following that, v0.2 improves the data delivery pipeline: instead of downloading the output of \textit{ServiceX}, in this version \textit{coffea} streams it directly out of the object storage to which \textit{ServiceX} writes its output.

Version v1.0 implements analysis task v1, handling the switch to input files in CMS NanoAOD format, with otherwise minimal changes to the analysis logic.
Subsequent releases in the v1 series fix minor bugs in the implementation and expand functionality for benchmarking purposes.

A version 2.0 implementation is planned to be included once the corresponding analysis task definition has been finalized.
The description of the implementation in these proceedings corresponds to a task version that already includes aspects of the upcoming version 2.0, in particular the \ac{ML} parts of the pipeline.

\section{Connections to analysis facilities and benchmarking}

The majority of \ac{AGC} reference implementation development and testing happens on analysis facilities deployed in the \textit{coffea-casa}~\cite{coffea-casa} model.
They provide the required software environment and services which are used in the \ac{AGC} reference implementation (such as \textit{ServiceX} and \textit{NVIDIA Triton}).
Users of \textit{coffea-casa} facilities are presented with a JupyterLab~\cite{jupyter} interface and can seamlessly scale to computing resource provide via a batch system by using \textit{Dask}.

Work on the \ac{AGC} project includes performance evaluations of \ac{AGC} implementations in order to identify and address bottlenecks.
First results are documented in proceedings from the ACAT 2022 conference~\cite{acat_proceedings}.
They demonstrate efficient scaling to hundreds of CPU cores for distributed execution and an efficient resource usage by scheduling with \textit{Dask}.

The \ac{AGC} reference implementation includes optional handles that allow altering the computational cost of the analysis task, as well as the amount of data that needs to be read from input files.
These allow probing additional types of workloads for benchmarking purposes, simulating analyses that are more likely to be I/O- or CPU-bound.

\section{Future plans}

The next major goal for the \ac{AGC} project is finalizing version v2 of the \ac{AGC} analysis task and providing a corresponding reference implementation.
It will feature a \ac{ML} component, which has already been implemented, as well as an extended set of systematic uncertainties to handle and correspondingly a larger amount of data to ingest and process.

A longer term goals is the implementation of a differentiable analysis pipeline to investigate the benefits of gradient-based automatic analysis optimization.
Another target for the \ac{AGC} is the demonstration of a "column joining" mechanism.
This refers to dynamically enhancing datasets used by physicists for analysis with missing information only available in parent datasets.
An example of this is automatically joining together event information provided by NanoAOD input files with additional information only found in MiniAOD datasets and presenting them to the user in a unified way.

\section{Conclusions}

The \ac{AGC} project develops and studies workflows for physics analysis at the HL-LHC.
It is centered around a physics analysis task and implementations thereof, which provide a central gathering point for the broader community and context for discussions about future analysis workflows.
The analysis task is based on using CMS Open Data to perform a \ttbar cross-section measurement, which has been extended to include a frequently requested \ac{ML} component.
IRIS-HEP provides a Python-based reference implementation for this analysis task.
All required data and the reference implementation are accessible via a GitHub repository.

\begin{acknowledgement}
     This work was supported by the U.S. National Science Foundation (NSF) Cooperative Agreement OAC-1836650 (IRIS-HEP).

     The \ac{AGC} is made possible thanks to the help of a large number of people working on many different projects.
     Thank you in particular to the teams behind: coffea-casa, Scikit-HEP, coffea, IRIS-HEP Analysis Systems, ServiceX, IRIS-HEP DOMA, IRIS-HEP SSL, and the CMS Data Preservation and Open Access (DPOA) group.
 \end{acknowledgement}

\bibliography{main}
\end{document}